\documentclass[epjST, final]{svjour} \usepackage{graphicx, subcaption,
  setspace, indentfirst, amsfonts, amsthm, empheq, color, soul}
\graphicspath{ {./} }
\newcommand\ignore[1]{}

\title{Equation-free analysis of a dynamically evolving multigraph}
\author{Alexander Holiday \inst{1} \and Ioannis G. Kevrekidis
  \inst{1,2}} \institute{Department of Chemical and Biological
  Engineering, Princeton University, Princeton, New Jersey 08544, USA
  \and PACM, Princeton University, Princeton, New Jersey 08544, USA}

\abstract{In order to illustrate the adaptation of traditional
  continuum numerical techniques to the study of complex network
  systems, we use the equation-free framework to analyze a dynamically
  evolving multigraph.
  This approach is based on coupling short intervals of direct dynamic
  network simulation with appropriately-defined lifting and
  restriction operators, mapping the detailed network description to
  suitable macroscopic (coarse-grained) variables and back.
  This enables the acceleration of direct simulations through Coarse
  Projective Integration (CPI), as well as the identification of
  coarse stationary states via a Newton-GMRES method.
  We also demonstrate the use of data-mining, both linear (principal
  component analysis, PCA) and nonlinear (diffusion maps, DMAPS) to
  determine good macroscopic variables (observables) through which one
  can coarse-grain the model.
  These results suggest methods for decreasing simulation times of
  dynamic real-world systems such as epidemiological network
  models. Additionally, the data-mining techniques could be applied to
  a diverse class of problems to search for a succint, low-dimensional
  description of the system in a small number of variables.}

\begin{document}
\maketitle
\begin{onehalfspace}

  \pagebreak

  \section{Introduction}
  \label{sec:intro}

  Over the past decades, complex networks have been used as the model
  of choice for a truly broad class of physical problems, ranging from
  highway traffic \cite{joubert_large-scale_2010} to brain
  connectivity \cite{hermundstad_learning_2011}.
  When modeling such systems, dynamics are typically defined at a very
  detailed, ``fine'' scale, specifying individual node and edge
  interactions; explicit closed equations governing network
  macroscopic (collective) properties are often unavailable
  \cite{durrett_graph_2012,joubert_large-scale_2010,roche_agent-based_2011,swaminathan_modeling_1998}.
  The detailed interactions are often complicated functions dependent
  on multiple system parameters, which, when one also accounts for
  large network sizes inherent in many interesting problems, can make
  systematic numerical exploration computationally prohibitive.
  The lack of \textit{explicit} macroscopic equations prohibits the
  use of traditional numerical techniques such as fixed-point
  computation and stability analysis, that could offer valuable
  insight into the network's behavior, leaving little alternative but
  to work with full direct simulations of the entire network.
  Faced with these inherent limitations, investigators must either
  restrict their attention to a more modest parameter space
  \cite{hodgkin_quantitative_1952} or simplify the network model,
  potentially removing important features
  \cite{brown_variability_1999}. \par

  Equation-free modeling offers a promise to circumvent these
  challenges, allowing one to investigate the complex network at a
  macroscopic level while retaining the effects of full system details
  \cite{kevrekidis_equation-free:_2004,gear_equation-free_2003}.
  Underlying this method is the assumption that, although we cannot
  analytically derive equations governing network evolution, such
  closed equations do, in principle, exist.
  Furthermore, these unavailable equations are assumed to involve a
  small number of dominant collective (coarse-grained) variables.
  The important features of the complete network can be in fact, by
  this assumption, well represented by these select observable
  quantities.
  This may seem too restrictive, yet it is exactly the behavior
  witnessed across many network types: despite the initial complexity
  of the full system configuration, certain collective network
  properties appear to evolve smoothly in time, while the evolution of
  other, ``secondary'' properties, can be strongly correlated with
  that of the few ``primary'' variables
  \cite{bold_equation-free_2014,rajendran_coarse_2011,siettos_equation-free_2011}.
  Once these significant variables are uncovered, we can combine short
  intervals of full system simulation with operators that map the full
  system description to and from its representative coarse variable
  ``summary'', thus enabling previously-infeasible system-level
  analysis (see Section \ref{sec:ef} for further details). \par

  Here, we apply this framework to a dynamically evolving multigraph
  model.
  This offers a test of the methodology in the previously unexplored
  context of multigraphs.
  We demonstrate the acceleration of direct network simulations
  through Coarse Projective Integration (CPI), and the location of
  coarse stationary states through a matrix-free Newton-GMRES method.
  In addition, principal component analysis (PCA) and diffusion maps
  (DMAPS), two well established dimensionality-reduction techniques,
  are shown to enable an algorithm for characterization of the
  underlying low-dimensional behavior of the system. \par

  The paper is organized as follows: we begin in Section \ref{sec:m}
  with a description of the multigraph dynamic model.
  Sections \ref{sec:ef} and \ref{sec:ng} provide most details of the
  equation-free approach, specify how it was applied to our system,
  and present subsequent results.
  Section \ref{sec:dr} summarizes our use of PCA and DMAPS, and
  assesses their effectiveness in analyzing hidden, low-dimensional
  structure in this network system.

  \section{Model description}
  \label{sec:m}

\begin{figure}[ht!]
    \centering
    \includegraphics[width=0.9\textwidth]{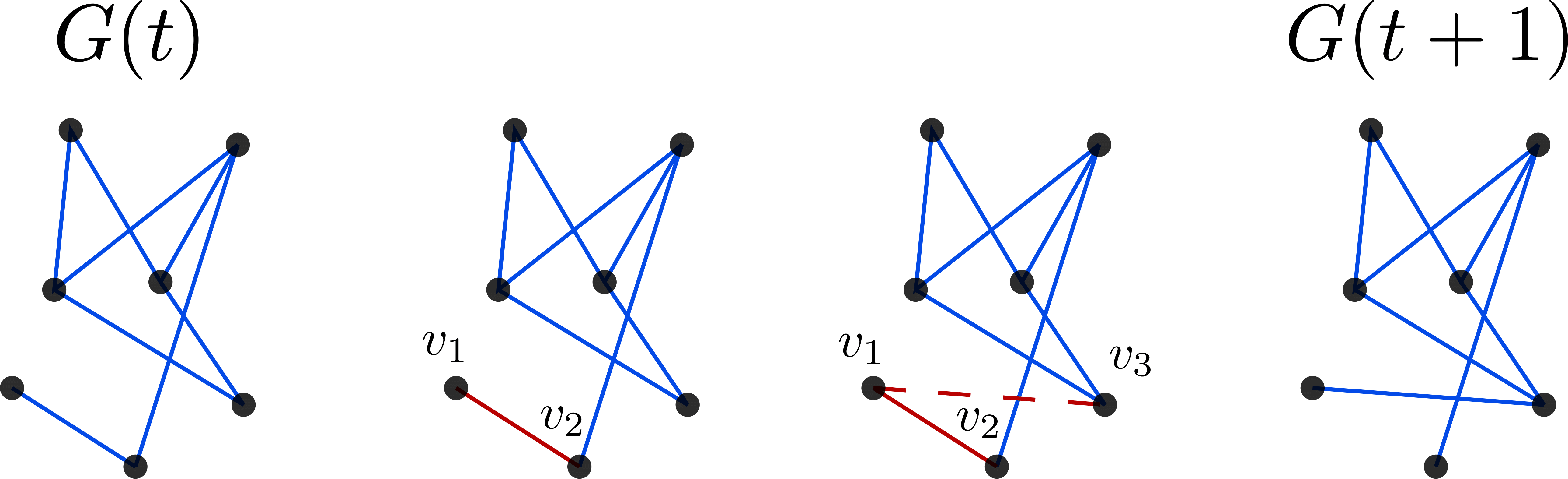}
    \caption{Schematic of the substeps of the evolution dynamics of the
      multigraph $G(t)$. \label{fig:step-illustration}}
  \end{figure}

  We study the edge-conservative preferential attachment model, a
  detailed mathematical analysis of which can be found in
  \cite{rath_time_2012} and \cite{rath_multigraph_2012}.
  The system evolves in discrete steps $t = 0,1,\ldots t_f$, and we
  denote the $n$-vertex graph at each point by $G_n(t)$.
  The initial state, $G_n(0)$, is an arbitrary distribution of $m$
  edges among the $n$ vertices; the total number of both edges and
  vertices is conserved.
  No restrictions are placed on this initial distribution: multiple
  edges and loops are permitted. 
  The system is then advanced step-by-step based on the following
  procedure:

  \begin{enumerate}
  \item Choose an edge $e_{ij} \in E(G)$ uniformly at random, and flip
    a coin to label one of the ends as $v_{i}$
  \item Choose a vertex $v_{k}$ using linear preferential attachment:
    $P(v_{k} = v_{l}) = \frac{d_{l} + \kappa}{\sum\limits_{i=1}^{n}
      d_{i} + n \kappa}$
    \label{step:prob}
  \item Replace $e_{ij}$ with $e_{ik}$,
  \end{enumerate}

  \noindent where $d_i$ is the degree of vertex $v_i$, $E(G)$ is the
  set of edges in the graph, and $\kappa \in (0, \infty)$ is a model
  parameter affecting the influence degrees have on the probability
  distribution in Step \ref{step:prob}. 
  That is, taking $\lim\limits{\kappa \rightarrow 0}$, we recover
  ``pure'' preferential attachment, and probability scales directly
  with degree,
  while
  $\lim\limits_{\kappa \rightarrow \infty} P(v_k = v_l) = \frac{1}{n}
  \; \forall \; l$, and individual degrees have no effect. A single
  step of this evolution process is illustrated in
  Fig. ({\ref{fig:step-illustration}}). Note that this can also be
  represented as a graph in which only one edge is permitted between
  each pair of nodes, but with edge weights signifying the number, or
  strength, of connections between nodes.
  \par

\begin{figure}[th!]
    \vspace{-5mm} \centering
    \begin{subfigure}[t]{0.49\textwidth}
      \centering
      \includegraphics[width=\textwidth]{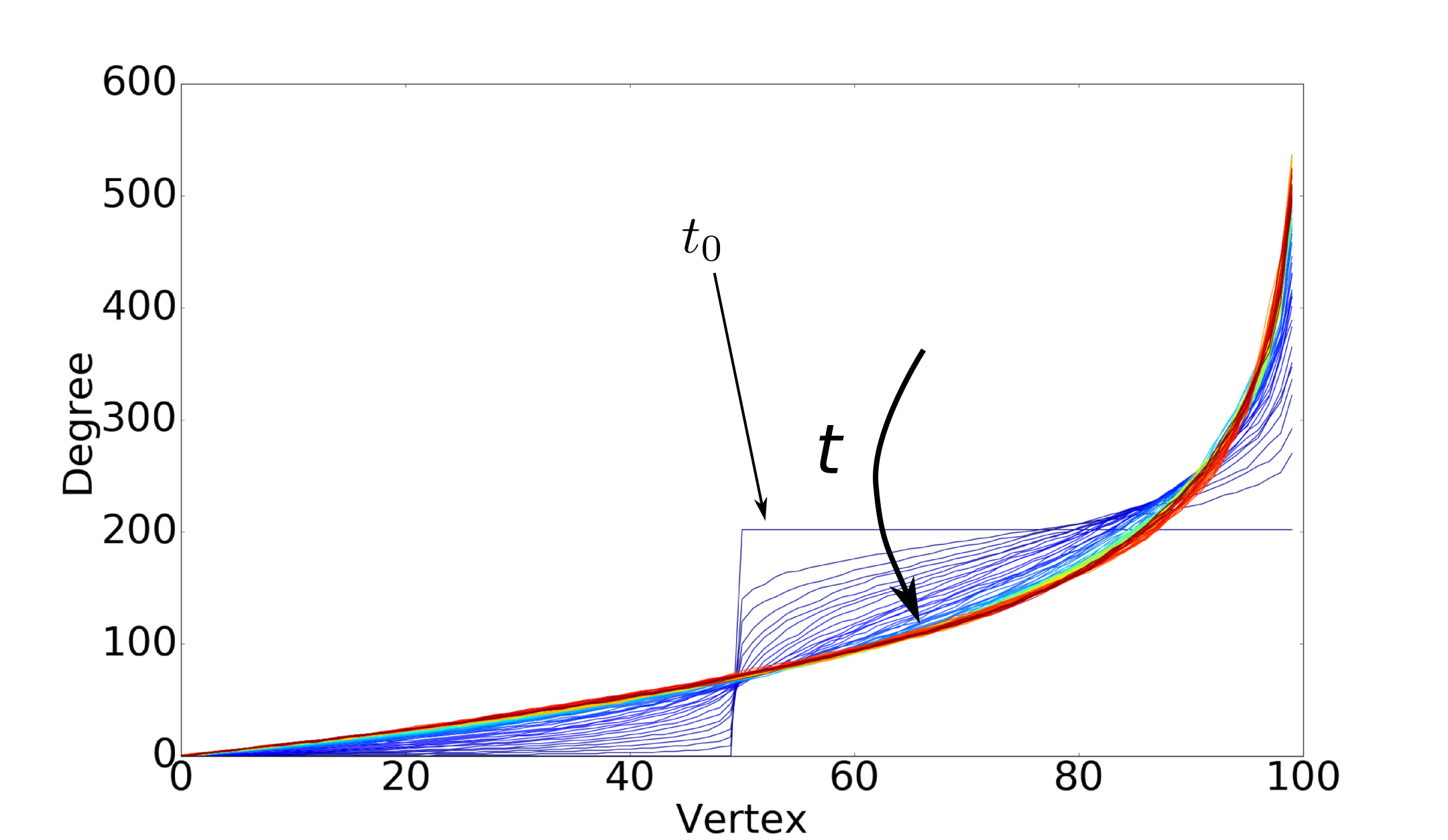}
      \subcaption{\label{fig:lopsided-init}}
    \end{subfigure} %
    \begin{subfigure}[t]{0.49\textwidth}
      \centering
      \includegraphics[width=\textwidth]{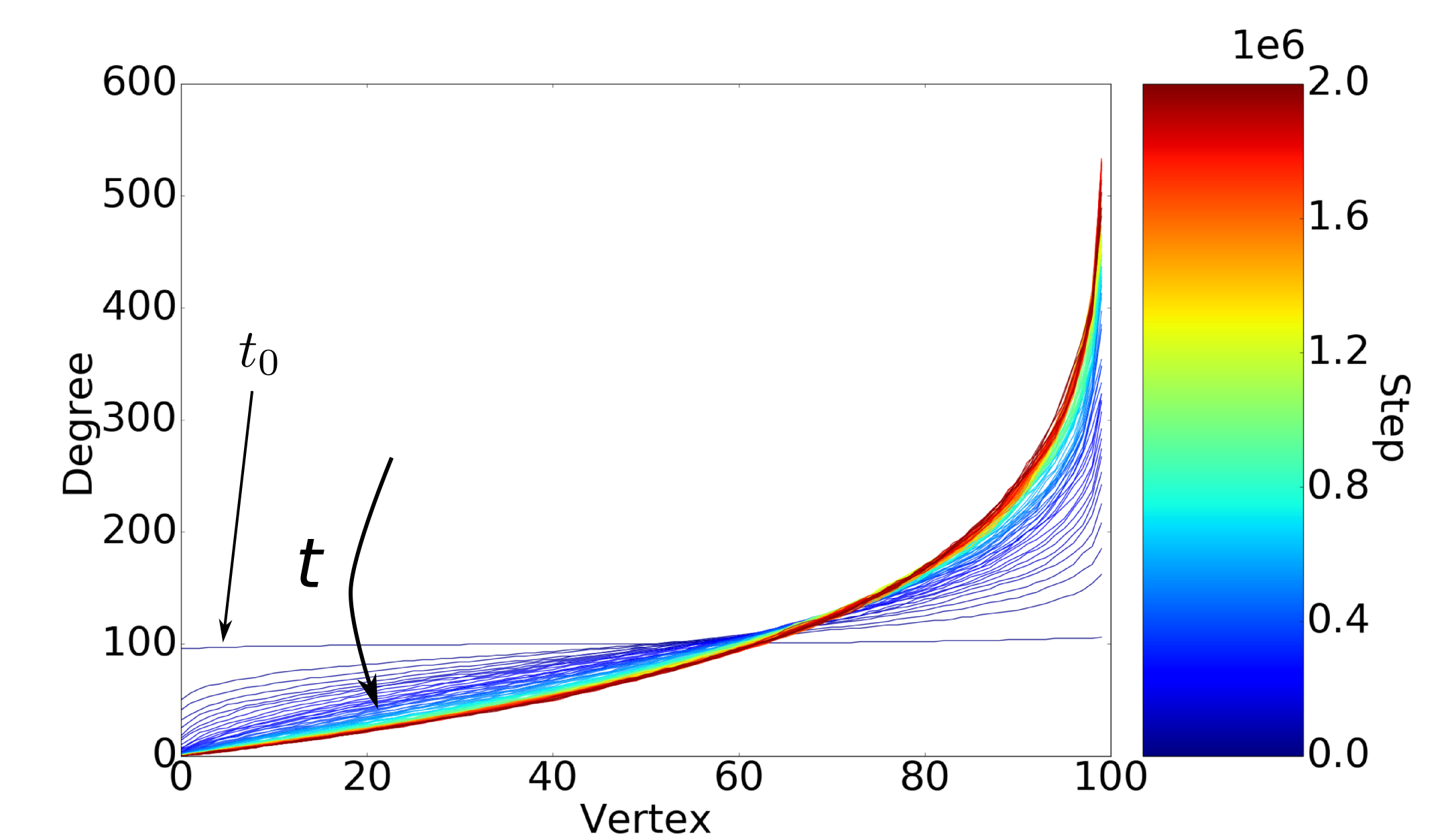}
      \subcaption{\label{fig:erdos-init}}
    \end{subfigure}
    \caption{Evolution of the multigraph model's sorted degree
      sequence. Two distinct transients are shown, initialized with
      (a) an Erd\H{o}s-R\'{e}nyi random graph with $m = 5,050$ edges
      and (b) a graph in which half the vertices are isolated, and
      half uniformly share $m = 5,050$ edges. Both approach the same
      ultimate stationary sequence. To smoothen the evolution of this
      stochastic system, here we plot the instantaneous average of twenty
      simulations. \label{fig:dse}}
  \end{figure}

  Evolving the system in this manner, the degree sequence approaches a
  stationary distribution over $O(n^3)$ steps.
  As explained in \cite{rath_time_2012}, this distribution is
  dependent only on the system parameters $\rho = \frac{2*m}{n}$ and
  $\kappa$.
  Fig. (\ref{fig:dse}) illustrates the evolution of the degree
  sequence of two networks with different initial configurations but
  the same values of $\rho$ and $\kappa$ respectively; as expected, we
  observe they approach an identical stationary state.

  \begin{figure}[th!]
    \centering
    \includegraphics[width=0.8\textwidth]{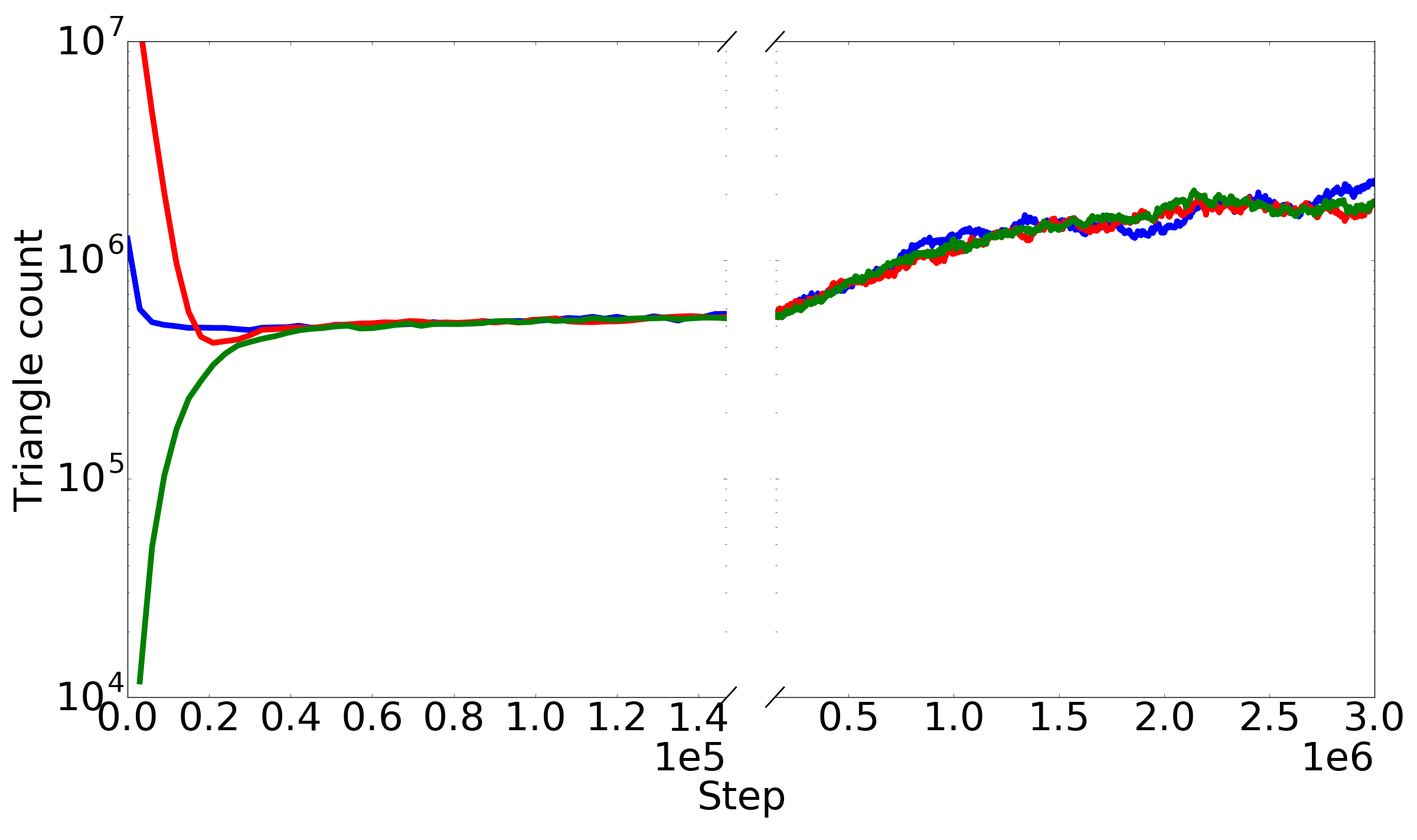}
    \caption{Evolution of higher-order network statistics (here, the
      total triangle count) for three different network
      initializations. They are quickly ($O(n^2)$ steps) drawn to a slow
      manifold on which they slowly evolve over ($O(n^3)$ steps) to a
      stationary state. \label{fig:sv}}
  \end{figure}

  \section{Equation-free modeling}
  \label{sec:ef}

  \subsection{Coarse-graining}

  A prerequisite to the implementation of equation-free algorithms is
  the determination of those few, select variables with which one can
  ``close'' a useful approximate description of the coarse-grained
  dynamics of the full, detailed network.
  This set of variables should clearly be (much) less than those of the
  full system state, and they should evolve smoothly (so, either our
  network would have a large number of nodes, or we would take
  expectations over several realizations of networks with the same
  collective features).
  Based on the profiles depicted in Fig. (\ref{fig:dse}), the graph's
  degree sequence, (equally informatively, its degree distribution) is a
  good candidate collective observable.
  It appears to evolve smoothly, while also providing significant
  savings in dimensionality, from an $O(n^2)$ adjacency matrix to a
  length-$n$ vector of degrees. \par

  It now becomes crucial to test whether the evolution of other
  properties of the graph can be correlated to the degree sequence.
  If not, our current description is incomplete (an equation cannot
  close with only the degree sequence) and there exist other important
  coarse variables that must be included in our macroscopic system
  description.
  To assess this, we constructed graphs with identical degree sequences
  but varying triangle counts and recorded the evolution of this
  observable under the dynamics prescribed in Section
  \ref{sec:m}. Fig. (\ref{fig:sv}) shows that, within a short time, the
  triangle counts are drawn to an apparent shared ``slow manifold'',
  despite the variation in initial conditions.
  This supports our selection of the degree sequence as an adequate
  coarse variable to model the system. \par

  Next we describe the other two key elements to equation-free modeling
  which map to and from our microscopic and macroscopic descriptions:
  restriction and lifting operators.

  \begin{figure}[ht!]
    \centering
    \includegraphics[width=\textwidth]{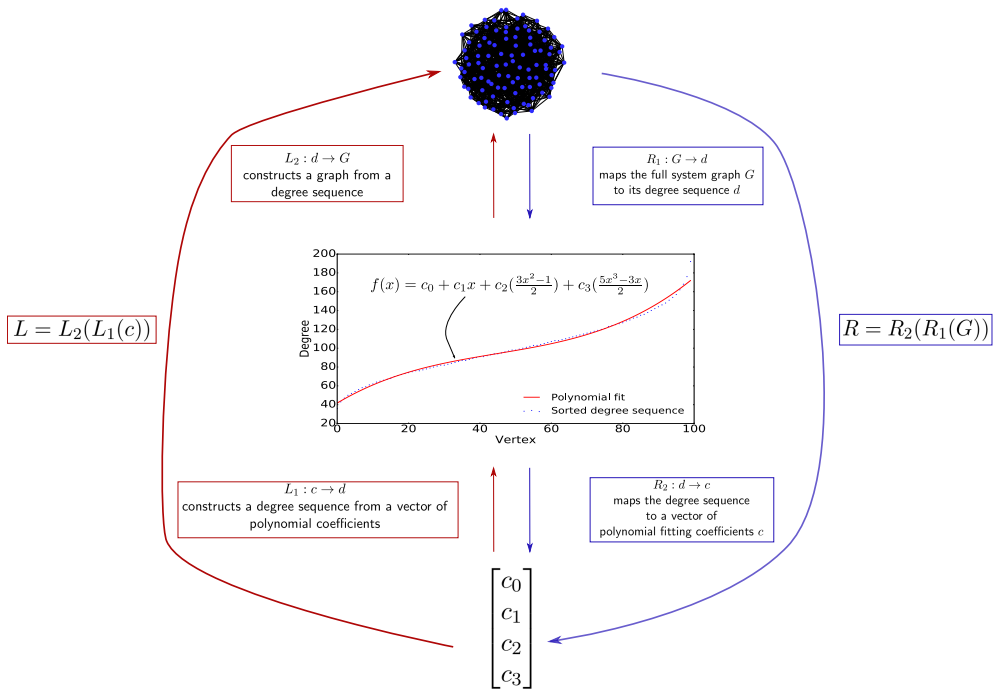}
    \caption{Schematic of our restriction ($\mathbf{R}$) and lifting
      ($\mathbf{L}$) procedures. The graph's sorted degree sequence is
      calculated ($R_1$) and then fit to a third-order polynomial
      ($R_2$). This maps graphs ($G$) to vectors of four coefficients
      ($\mathbf{c}$). To lift, the polynomial is used to generate a
      degree sequence ($L_1$), which is then used to construct a
      consistent graph ($L_2$). This process maps vectors of four
      coefficients ($\mathbf{c}$) to graphs ($G$).
      $\mathbf{R} = R_2 \circ R_1 $ is our restriction and
      $\mathbf{L} = L_2 \circ L_1$ is our
      lifting. \label{fig:cpi-diagram}}
  \end{figure}

  \subsection{Restriction}

  The restriction operator ``translates'' the full network description
  to its corresponding coarse variable values.
  Here, this involves mapping a multigraph to its sorted degree
  sequence, a simple calculation.
  However, we may be able to further reduce the dimensionality of our
  coarse system by keeping not the entire length-$n$ degree sequence,
  but rather a low-order polynomial fitting of it.
  To do so, we first sort the sequence to achieve a smooth,
  monotonically increasing dataset, then fit the result to a third-order
  polynomial, which had been observed to closely approximate the
  sequence evolution throughout time.
  Our coarse description thus consisted, at every moment in time, of the
  four polynomial coefficient values, specifying a particular sorted
  degree sequence, as shown in Fig. (\ref{fig:cpi-diagram}).
  The restriction operator therefore maps multigraphs to length-four
  coefficient vectors.
  This procedure is represented by the blue arrows of
  Fig. (\ref{fig:cpi-diagram}).

  \subsection{Lifting}

  The lifting operator performs the inverse role of restriction:
  translating coarse variables into full networks.
  This is clearly a one-to-many mapping.
  Specifically in the context of our model, we map vectors of polynomial
  coefficients to full networks in a two-stage process.
  First, the coefficient vector is used to recreate a sorted degree
  sequence by evaluating the full polynomial at integer values and
  rounding to the nearest degree, as depicted in
  Fig. (\ref{fig:cpi-diagram}).
  If the sum of the resulting degree sequence is odd, a single degree is
  added to the largest value to ensure the sequence is graphical.
  Second, this degree sequence is used as input to any algorithm that
  produces a network consistent with a graphical degree sequence (here
  we typically used a Havel-Hakimi algorithm, creating a graph whose
  degree sequence matches that of the input
  \cite{havel_remark_1955,hakimi_realizability_1962}).
  While the canonical Havel-Hakimi method produces simple graphs, it is
  not difficult to extend it to multigraphs by allowing the algorithm to
  wrap around the sequence, producing multiple edges and self-loops.
  The lifting procedure is represented by the red arrows of
  Fig. (\ref{fig:cpi-diagram}).

  \subsection{Coarse projective integration (CPI)}
  \label{sec:cpi}

  The components described above are combined to accelerate simulations
  through Coarse Projective Integration.
  Denoting the lifting operator by $\mathbf{L} \; : \; c \rightarrow G$
  where $c \in \mathbb{R}^4$ is the vector of polynomial coefficients,
  and the restriction operator as $\mathbf{R} \; : G \rightarrow c$, the
  algorithm progresses as follows:

  \begin{enumerate}
  \item Advance the full system for $t_h$ steps
    \label{cpi:heal}
  \item Continue for $t_c$ steps, restricting the network at intervals
    and recording the coarse variable values
  \item Using the coarse variables collected over the previous $t_c$
    steps, project each variable forward $t_p$ steps with, here, a
    forward-Euler method
    \label{cpi:proj}
  \item With the new, projected coarse variables, lift to one (or more
    copies of) full network
    \label{cpi:init}
  \item Repeat from Step (\ref{cpi:heal}) until the desired time has
    been reached.
  \end{enumerate}

  Note that Step (\ref{cpi:heal}) is intended to allow a singularly
  perturbed quantity to approach its slow manifold.
  Upon initializing a new full system (here, a new network) in Step
  (\ref{cpi:init}), higher order observables (for example, the triangle
  count) will be far from the values they would attain in a detailed
  simulation.
  If the coarse system model closes with our chosen variables, then
  either (a) these quantities do not affect the dynamics of the degree
  sequence; or (b) they do, but then by hypothesis these quantities
  quickly evolve to functions of the selected coarse variables (i.e.,
  they approach the slow manifold).
  In the second case, after a short interval of ``healing'' they will be
  drawn to the expected trajectory, after which we begin sampling.
  This is analogous to Fig. (\ref{fig:sv}).
  The computational gains arise from the projective step,
  (\ref{cpi:proj}).
  Here, we advance the system $t_p$ steps at the cost of one evaluation of forward Euler, instead of $t_p$ direct detailed steps. \\

\begin{figure}[h!]
    \vspace{-5mm} \centering
    \begin{subfigure}{0.59\textwidth}
      \centering
      \includegraphics[width=\textwidth]{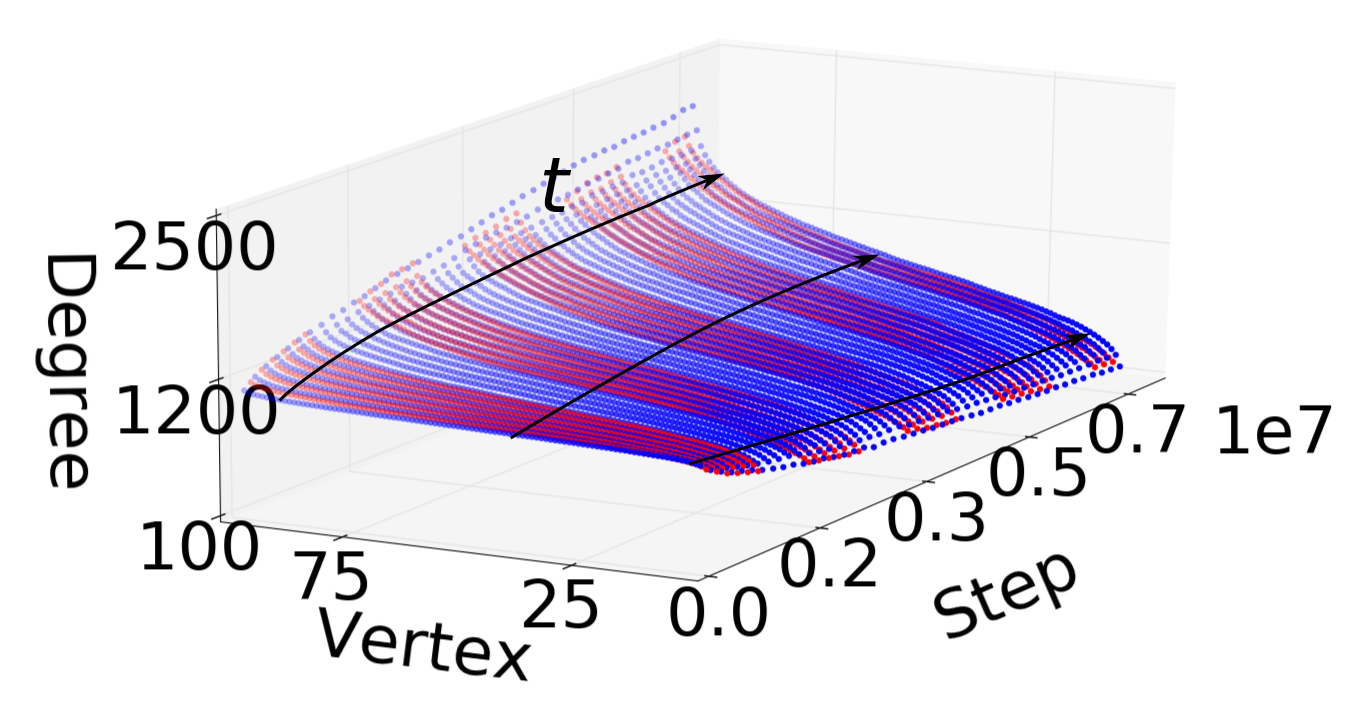}
      \subcaption{\label{fig:cpi-error}}
    \end{subfigure} %
    \begin{subfigure}{0.39\textwidth}
      \centering
      \includegraphics[width=\textwidth]{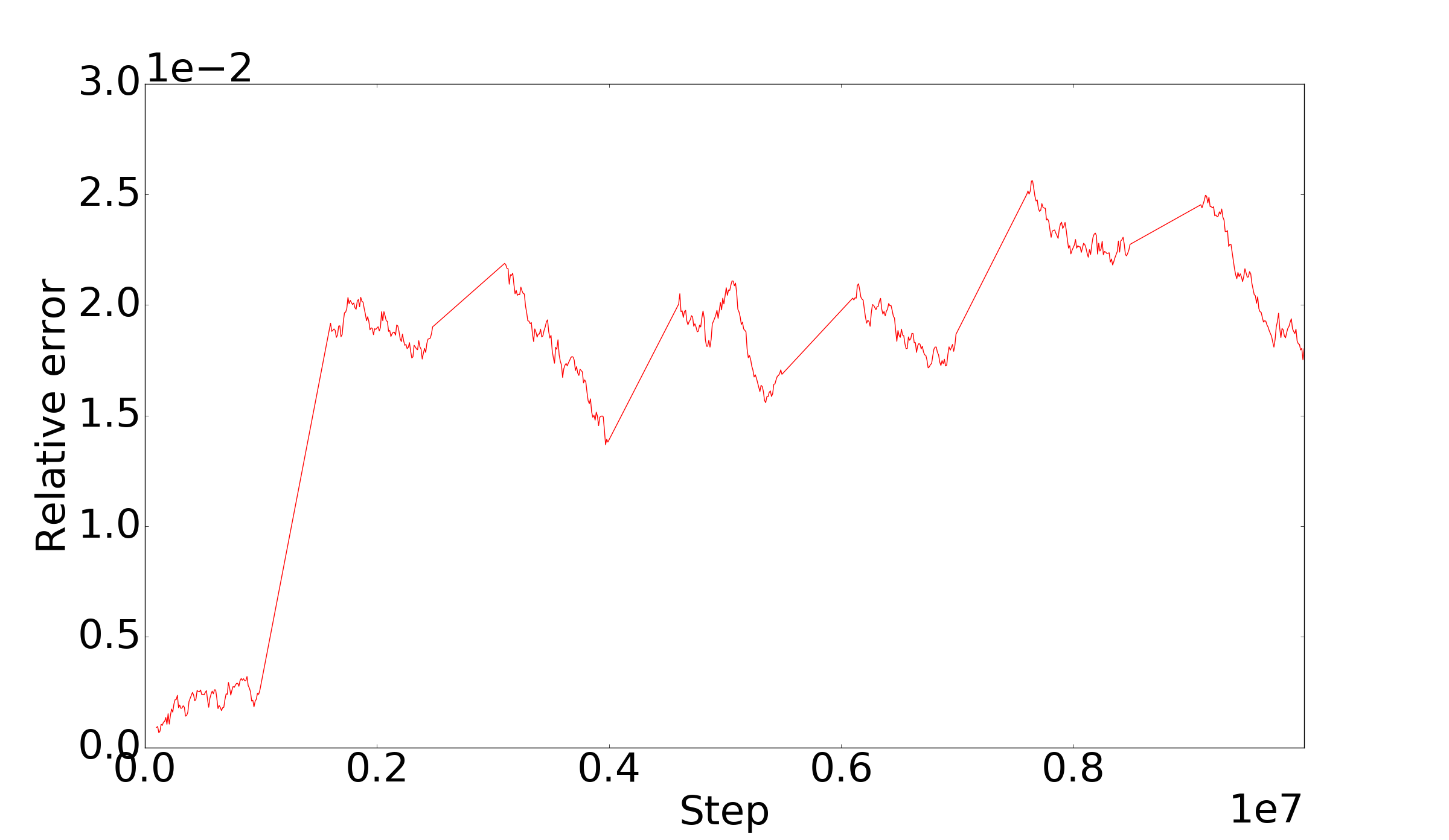}
      \subcaption{\label{fig:self-error}}
    \end{subfigure}%
    \caption{CPI results: (a) the evolution of the CPI-accelerated
      degree sequence (red) compared to direct simulation (blue) and (b)
      error in CPI-accelerated runs calculated by comparing
      CPI-accelerated degree sequences to those arising from an ensemble
      of direct simulations. \label{fig:cpi-results}}
  \end{figure}

  Results of the application of this general framework with the specific
  lifting and restriction operators previously outlined are shown in
  Fig. (\ref{fig:cpi-results}). We used an $n=100$-vertex graph with
  $m=50000$ edges and parameter value $\kappa=1$. We ran the model for a
  total of $10 \cdot n^3$ steps, with $t_h = 10 \cdot n^2$, $t_c = n^3$
  and $t_p = 50 \cdot n^2$.
  We see good agreement between the CPI-accelerated and normal systems,
  while reducing the number of detailed steps by one third.
  It is important to mention that this method was applied not to a
  single system instantiation,
  but to an ensemble of fifty such realizations.
  This ensured that when properties such as the fitted polynomial
  coefficients were averaged over the ensemble they evolved smoothly
  despite the stochastic nature of the system; in effect, we are
  computing the ``expected network evolution'' averaged over consistent
  realizations.

  \section{Coarse Newton-GMRES}
  \label{sec:ng}
  Aside from the computational savings of model simulations offered by
  CPI, the equation free framework also permits the calculation of
  system steady states through fixed point (here, matrix-free
  Newton-GMRES) algorithms.
  Referring back to the CPI procedure outlined in Sec. (\ref{sec:cpi}),
  we can define an operator $\Theta: \mathbf{d} \rightarrow \mathbf{d}$
  projecting coarse variables at $t$ to their values at $t + \delta t$:
  $\mathbf{d}(t+\delta t) =\Theta(\mathbf{d}(t))$.
  Note that in this section, we take the sorted degree sequence as our
  coarse variable.
  A system fixed point could then be located by employing Newton's
  method to solve the equation

  \begin{align}
    \label{eq:f}
    \mathbf{F}(\mathbf{d}) \equiv \Theta(\mathbf{d}) - \mathbf{d} = 0
  \end{align}

  However, this requires the repeated solution of the system of linear
  equations
  $DF(\mathbf{d}^{(k)}) \cdot \delta \mathbf{d}^{(k+1)} =
  -\mathbf{F}(\mathbf{d}^{(j)})$,
  in which the system Jacobian, $DF$ is unavailable.
  Thankfully, we may circumvent this obstacle by estimating the
  directional derivatives $DF(\mathbf{d}) \cdot \delta \mathbf{d}$ via a
  difference approximation of the form

  \begin{align}
    DF(d) \cdot \delta \mathbf{d} \approx \frac{\| \delta \mathbf{d} \| \mathbf{F}(\mathbf{d} + h \| \mathbf{d} \| \frac{\delta \mathbf{d}}{\| \delta \mathbf{d} \|}) - \mathbf{F}(\mathbf{d})}{h \| \mathbf{d} \|}
  \end{align}

  \noindent for nonzero $\mathbf{d}$, which in turn is evaluated through
  calls to $\mathbf{F}$ as defined in Eq. (\ref{eq:f})

\begin{figure}[h!]
    \vspace{-5mm} \centering
    \begin{subfigure}{0.49\textwidth}
      \centering
      \includegraphics[width=\textwidth]{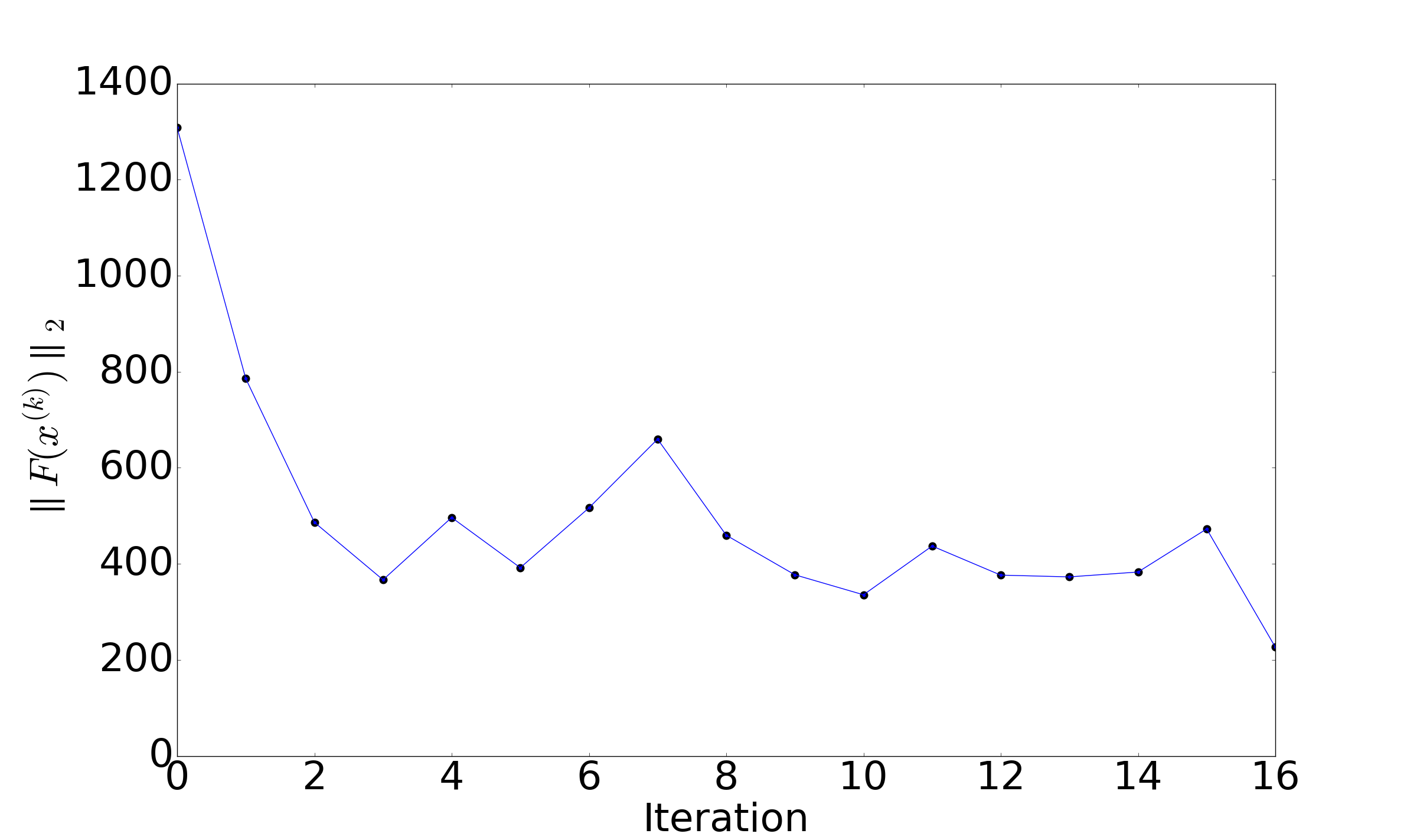}
      \subcaption{\label{fig:newton-error}}
    \end{subfigure} %
    \begin{subfigure}{0.49\textwidth}
      \centering
      \includegraphics[width=\textwidth]{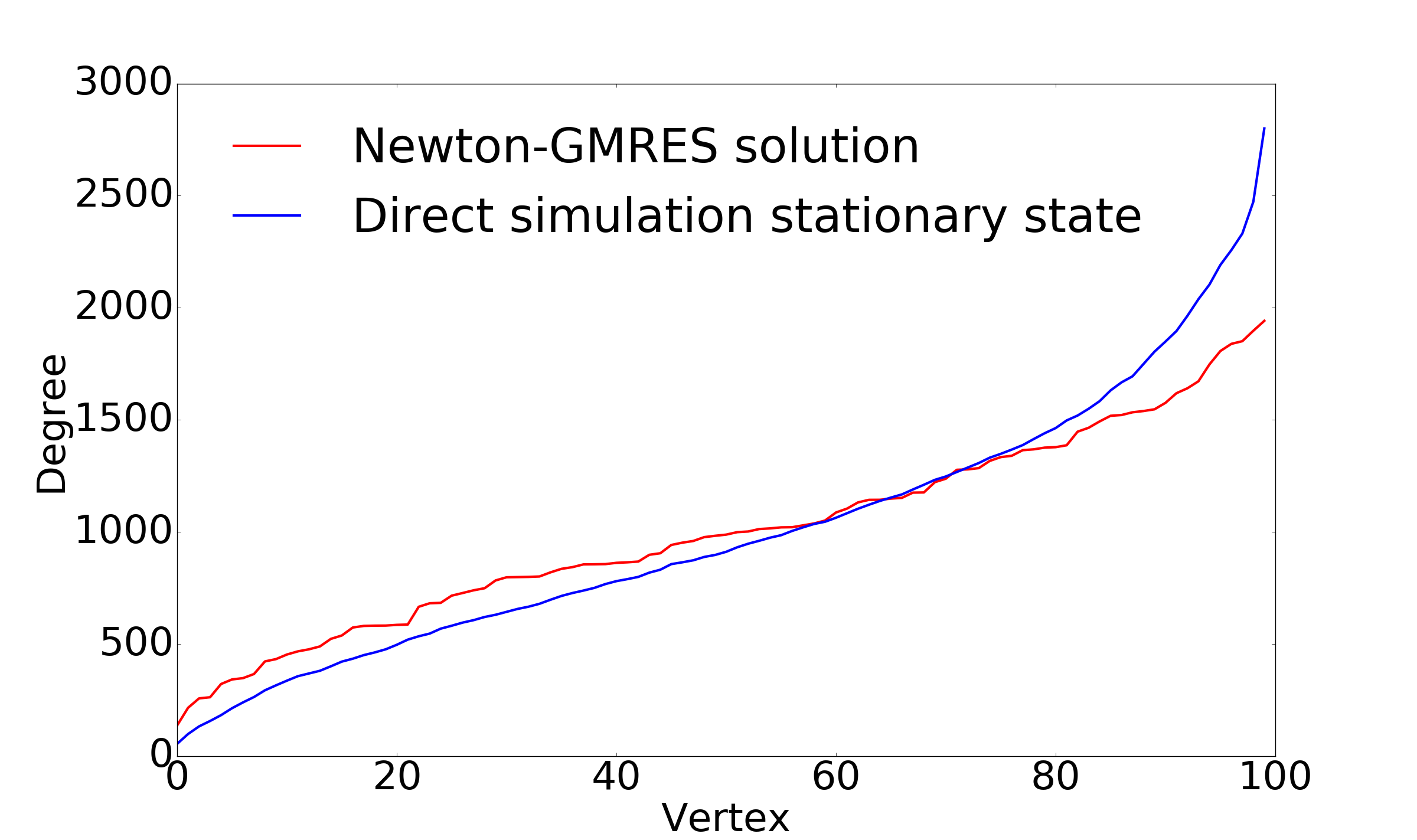}
      \subcaption{\label{fig:newton-comp}}
    \end{subfigure}%
    \caption{Coarse Newton-GMRES results: (a) evolution of the error in
      the coarse Newton-GMRES iteration scheme; and (b) visual
      comparison of the algorithm's solution to the stationary state
      obtained from direct simulation. \label{fig:newton-results}}
  \end{figure}

  This is precisely the approach of the Newton-GMRES method, in which
  the solution to a linear system is sought for in expanding Krylov
  subspaces \cite{kelley_solving_2003}.
  Applying this algorithm in conjunction with the $\Theta$ operator
  defined in Sec. (\ref{sec:cpi}) allowed us to locate the stationary
  distribution without simply running and observing the full system for
  long times, as would otherwise be required.
  Results are shown in Fig. (\ref{fig:newton-results}).
  We note that coarse stability results can also be obtained via an
  iterative Arnoldi eigensolver, again estimating matrix-vector products
  as above.

  \section{Algorithmic coarse-graining}
  \label{sec:dr}
  Crucial to the above analysis was the determination of suitable
  coarse, system variables: it is the starting point of any equation
  free method.
  However, the discovery of such a low-dimensional description is highly
  non-trivial.
  Currently, as in this paper, they are ``discovered through informed
  experience'': through careful investigation of direct simulations, and
  knowledge of previous analytical results.
  Clearly, any process which could algorithmically guide this search
  based on only simulation data would be of great benefit to the
  modeling practitioner.
  We now illustrate two such algorithms: principal component analysis
  (PCA) and diffusion maps (DMAPS).
  First, we briefly discuss some aspects of the important issue of
  defining distances between networks, a prerequisite for any
  dimensionality-reduction technique.

  \subsection{On network distances}

  When applying dimensionality reduction techniques to a dataset, it is
  necessary to define a distance (or similarity) between each pair of
  data points.
  If these points are a set of vectors in $\mathbb{R}^n$ one has a
  number of metrics to choose from, the Euclidean distance being a
  common first option.
  Unfortunately, when individual points are not vectors but networks,
  the definition of a useful and computationally easily quantified
  metric becomes far more challenging.
  Examples such as the maximal common subgraph and edit distances,
  defined in \cite{bunke_graph_1998} and \cite{gao_survey_2010} define
  metrics on the space of graphs, but their computation is
  $NP\textnormal{-}hard$.
  Other computationally feasible approaches include comparing
  distributions of random walks on each graph
  \cite{vishwanathan_graph_2010}, calculating the so-called $n$-tangle
  density \cite{gallos_revealing_2014}, or calculating the edit distance
  with an approximation algorithm \cite{riesen_approximate_2009,zeng_comparing_2009}.  \\

\begin{figure}[h!]
    \vspace{-5mm} \centering
    \includegraphics[width=\textwidth]{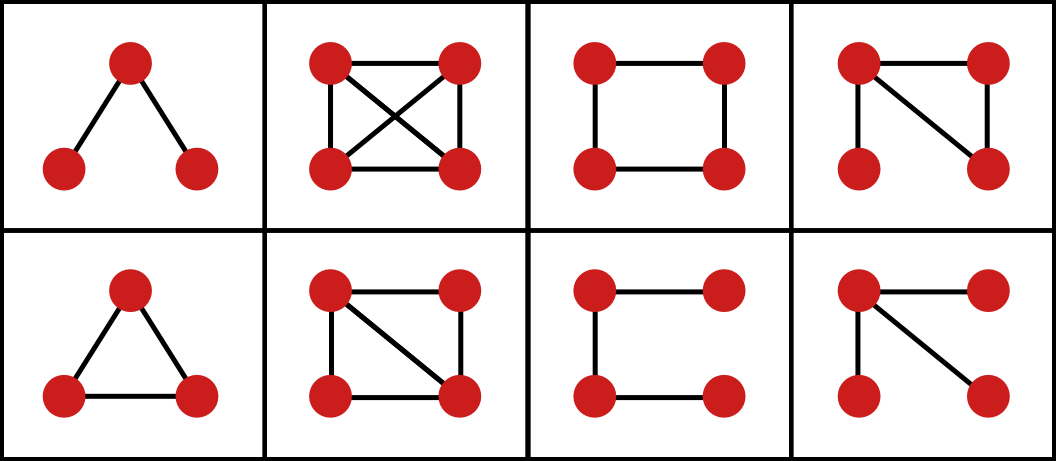}
    \caption{List of the single-edge subgraphs used to embed each
      network.  The number of times each appeared in the simplified
      input graph was calculated, mapping input graphs to
      $\mathbb{R}^8$. \label{fig:motifs}}
  \end{figure}

  The strategy used in the following computations, detailed in
  \cite{rajendran_analysis_2013} and \cite{xiao_structure-based_2008},
  enumerates the number of times a certain set of motifs (or subgraphs)
  appears in each network in the dataset.
  This maps each network to a vector in $\mathbb{R}^n$, and the
  Euclidean distance is subsequently used to quantify the similarity of
  two graphs.
  Due to computational restrictions, we chose here to only count the
  number of three- and four-vertex single-edge subgraphs contained in
  each network.
  As there are eight such motifs, shown in Fig. (\ref{fig:motifs}), this
  process $\gamma$ maps each graph to an eight-dimensional vector:
  $\gamma : G \rightarrow \mathbb{R}^8$.
  We applied this operation to a simplified version of each graph,
  wherein any multiple edges were reduced to a single edge.

  \subsection{PCA}
  \label{sec:pca}

  PCA is used to embed data into linear subspaces that capture the
  directions along which the data varies most
  \cite{jolliffe_principal_2014}.
  Given some matrix $X \in \mathbb{R}^{n \times m}$ in which each of the
  $m$ column vectors $x_i$ represents a different collection of the $n$
  system variables (i.e. a different data point), PCA computes a reduced
  set of $k<n$, orthogonal ``principal components''
  $z_i \in \mathbb{R}^n$ that constitute an optimal basis for the data
  in the sense that, among linear, $k$-dimensional embeddings,
  $w_i = Z^Tx_i$ captures the maximum possible variance in the dataset,
  where
  $Z = \begin{bmatrix} | & | & & | \\ z_1 & z_2 & \hdots & z_k \\ | & |
    & & | \end{bmatrix}$.
  This approach has found wide application, but can suffer from its
  inability to uncover simple \textit{nonlinear} relationships among the
  input variables.
  Indeed, many datasets will not ``just lie along some hyperplane''
  embedded in $\mathbb{R}^n$, but will rather lie on a low-dimensional
  nonlinear manifold throughout the space. \\

\begin{figure}[h!]
    \vspace{-5mm} \centering
    \begin{subfigure}{0.49\textwidth}
      \centering
      \includegraphics[width=\textwidth]{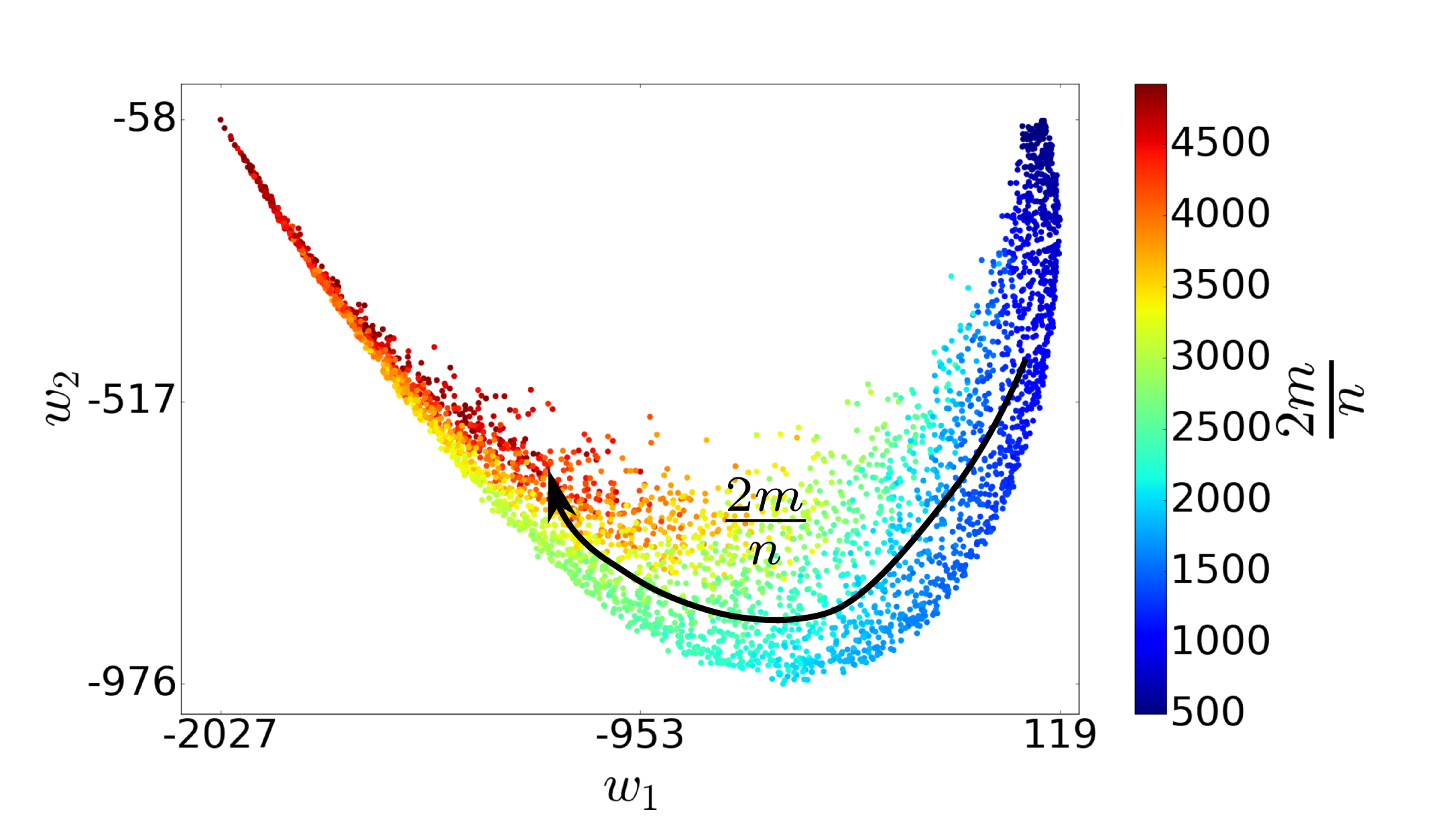}
      \subcaption{\label{fig:pca-rho}}
    \end{subfigure} %
    \begin{subfigure}{0.49\textwidth}
      \centering
      \includegraphics[width=\textwidth]{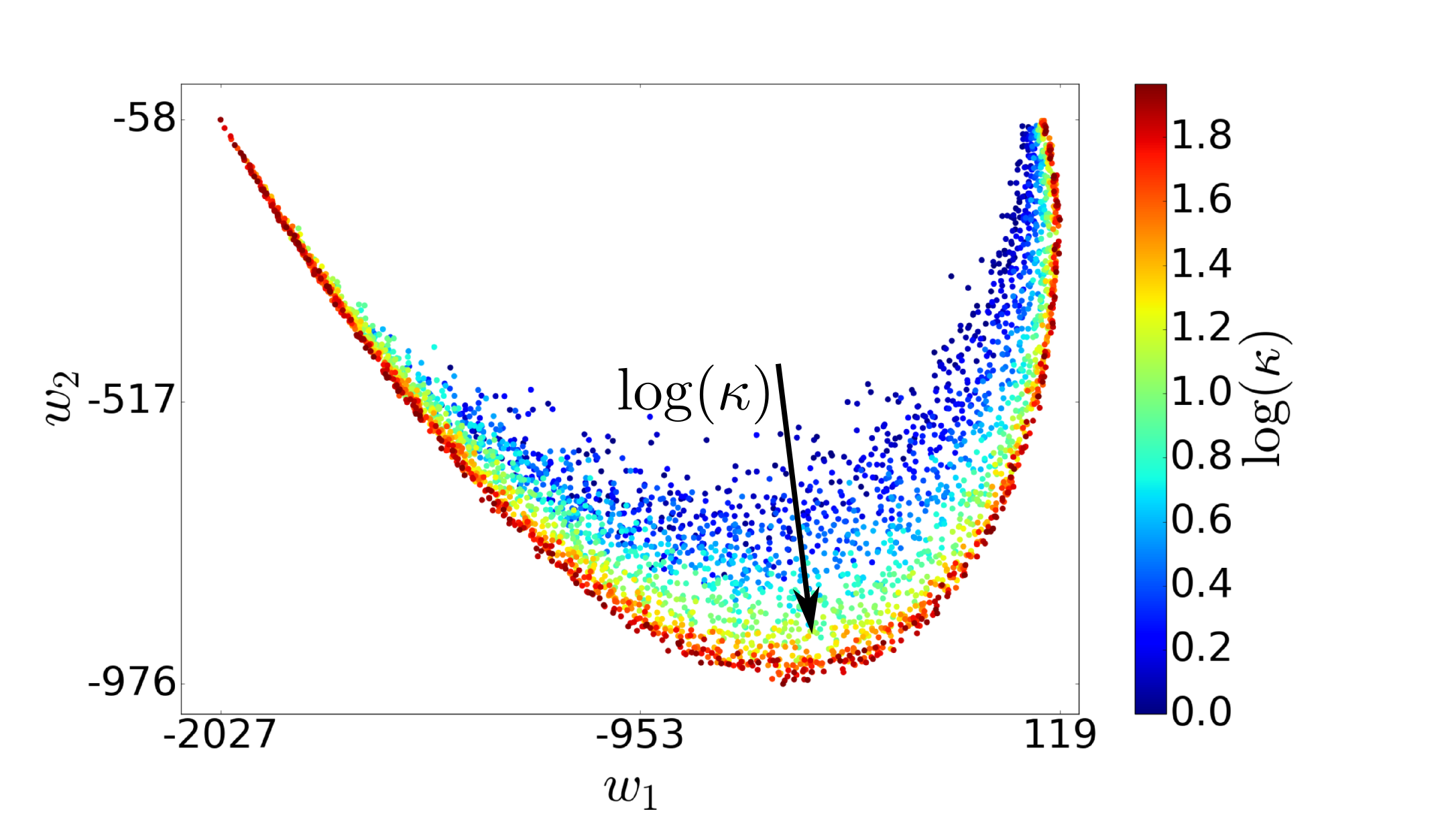}
      \subcaption{\label{fig:pca-kappa}}
    \end{subfigure}%
    \caption{PCA of motif-based embeddings: (a) coloring the
      two-dimensional PCA embedding with $\rho$ and (b) coloring the
      two-dimensional PCA embedding with $\kappa$. \label{fig:pca}}
  \end{figure}

  Theoretical results in \cite{rath_time_2012} state that, for a given
  network size $n$, the final stationary state depends only on the
  number of edges present $m$ and the model parameter $\kappa$.
  To assess both the validity of our graph embedding technique and the
  usefulness of PCA in the context of complex networks, we generated a
  dataset of stationary states over a range of $m \in [50, 5000]$ and
  $\log(\kappa) \in [0, 2]$ values by directly running the model over
  $2n^3$ steps ($N=30$ values of each parameter were taken, for a total
  of $900$ networks).
  We fixed the network size to $n=50$ vertices.
  Each resulting graph $G(m_i, \kappa_j) = G_{ij}$ was then embedded
  into $\mathbb{R}^8$ by counting the number of times each of the
  subgraphs shown in Fig. (\ref{fig:motifs}) appeared in the network.
  Thus $\gamma(G_{ij}) = v_{ij} \in \mathbb{R}^8$.
  We then proceeded to perform a principal component analysis on this
  collection of vectors $\{v_{ij}\}_{i,j=1}^N$.
  Interestingly, the first two principal components $z_1$ and $z_2$
  succeeded in uncovering a two-dimensional embedding of the dataset
  corresponding to the two underlying parameters $m$ and $\kappa$, as
  shown in Fig. (\ref{fig:pca}) in which the data is projected on the
  plane spanned by these two vectors.
  This suggests that, given some final network state $G(m, \kappa)$, by
  projecting its embedding onto these first two principal components one
  could obtain a reasonable approximation of the hidden parameter
  $\kappa$.

  \subsection{Diffusion Maps}

  Unlike PCA, DMAPS uncovers parameterizations of nonlinear manifolds
  hidden in the data.
  This is achieved by solving for the discrete equivalent of the
  eigenfunctions and eigenvalues of the Laplace-Beltrami operator over
  the manifold, which amounts to calculating leading
  eigenvector/eigenvalue pairs of a Markov matrix $A$ describing a
  diffusion process on the dataset.
  As the eigenfunctions of the Laplace-Beltrami operator provide
  parameterizations of the underlying domain, the output eigenvectors
  $\Phi_i$ from DMAPS similarly reveal and parameterize any hidden
  nonlinear structure in the input dataset. The $k$-dimensional DMAP of
  point $x_i$ is given by

  \begin{align*}
    \Psi(x_i; t) = \begin{bmatrix} \lambda_1^t \Phi_1(i) \\ \lambda_2^t
      \Phi_2(i) \\ \vdots \\
      \lambda_k^t \Phi_k(i) \end{bmatrix}
  \end{align*}

  \noindent where $\lambda_i$ is the $i^{th}$ eigenvalue of the Markov
  matrix $A$, $\Phi_i(j)$ the $j^{th}$ entry of eigenvector $i$, and the
  parameter $t$ allows one to probe multiscale features in the
  dataset. See \cite{coifman_diffusion_2006,nadler_diffusion_2006} for
  further details. \\

\begin{figure}[h!]
    \vspace{-5mm} \centering
    \begin{subfigure}{0.49\textwidth}
      \centering
      \includegraphics[width=\textwidth]{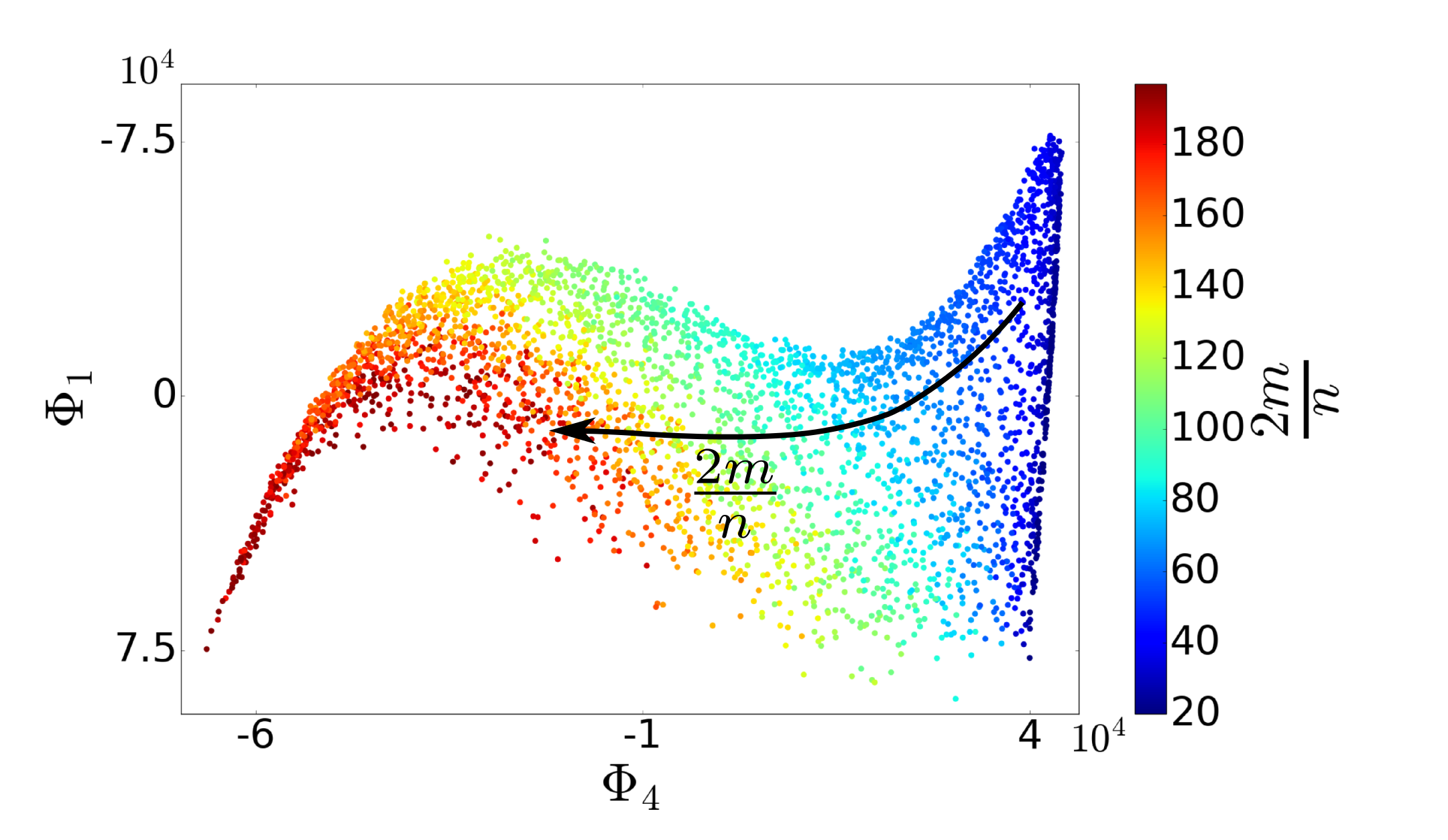}
      \subcaption{\label{fig:dmaps-rho}}
    \end{subfigure} %
    \begin{subfigure}{0.49\textwidth}
      \centering
      \includegraphics[width=\textwidth]{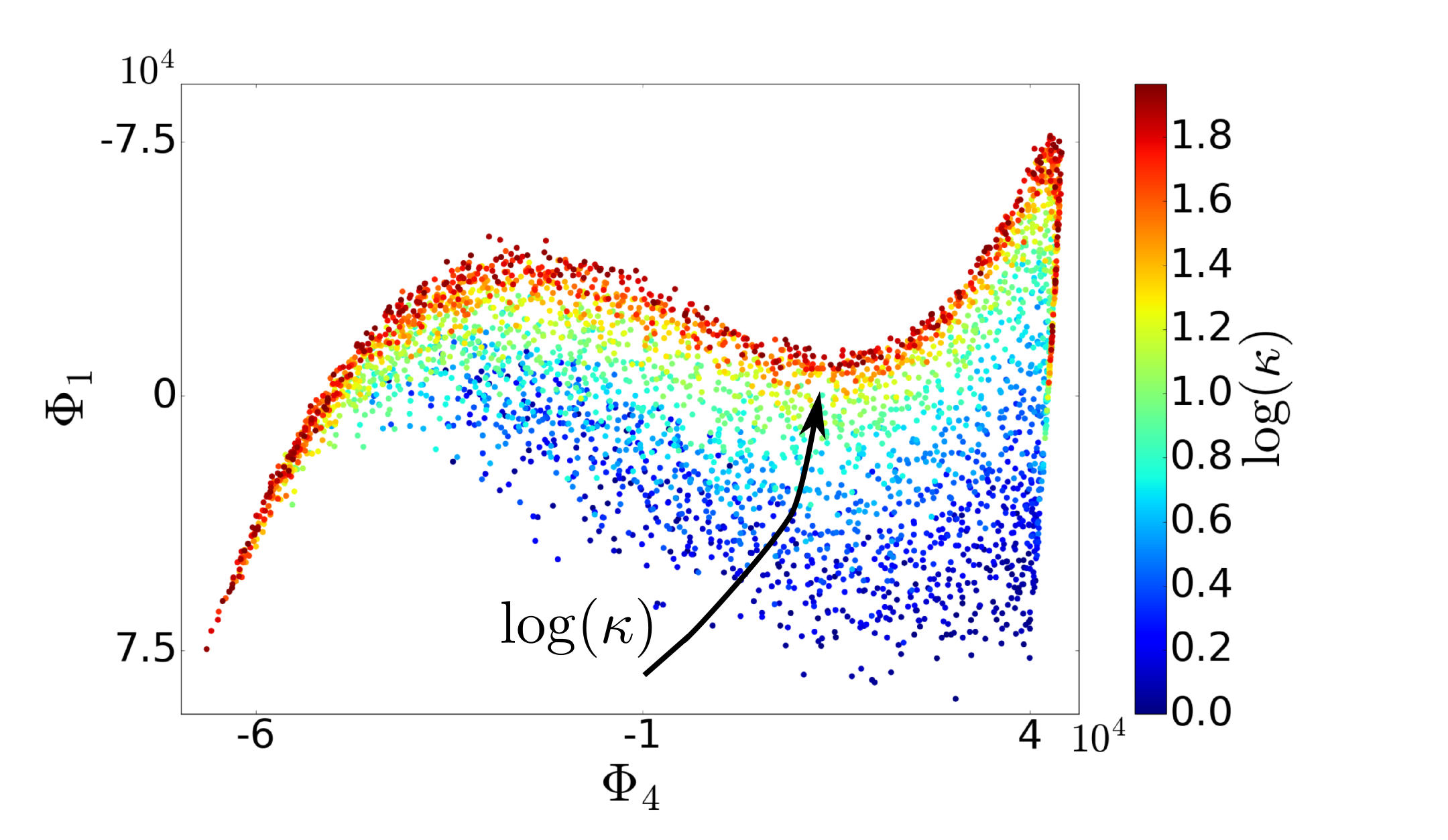}
      \subcaption{\label{fig:dmaps-kappa}}
    \end{subfigure}%
    \caption{DMAP of motif-based embeddings from a collection of
      simulations run at different parameter values: (a) coloring the
      two-dimensional DMAPS embedding with $\rho$ and (b) coloring the
      two-dimensional DMAPS embedding with $\kappa$. As with PCA, DMAPS
      uncovered the parameters governing the stationary state, known
      from \cite{rath_time_2012} \label{fig:dmaps-rk}}
  \end{figure}

  First we applied DMAPS to the dataset described in
  Sec. (\ref{sec:pca}).
  Given the apparent success of PCA in this setting, one would expect
  DMAPS to also uncover the two-dimensional embedding corresponding to
  different values of $m$ and $\kappa$.
  Fig. (\ref{fig:dmaps-rk}) shows that this is indeed the case: using
  $\Phi_1$ and $\Phi_4$ to embed the graphs produces a two-dimensional
  surface along which both $\rho$ and $\kappa$ vary independently. \\

  Additionally, we were interested in embeddings of different model
  trajectories.
  This dataset was generated by sampling two different model simulations
  as they evolved ($N=2000$ points were sampled from each trajectory).
  The parameters $n$, $m$ and $\kappa$ were held constant at $200$,
  $20100$ and $1.0$ respectively, but one graph was initialized as an
  Erd\H{o}s-R\'{e}nyi random graph (Fig. (\ref{fig:erdos-init})), while
  the other was initialized as a ``lopsided'' graph
  (Fig. (\ref{fig:lopsided-init})).
  Every $500$ steps the graph would be recorded till $N$ snapshots were
  taken of each trajectory, for a total of $1000000$ steps.
  Letting $G_e(t)$ refer to the Erd\H{o}s-R\'{e}nyi-initialized system
  at step $t$, and $G_l(t)$ the lopsided-initialized system, the
  embedding $\gamma$ was used to create points
  $\gamma(G_e(t)) = v_e(t) \in \mathbb{R}^8$ and
  similarly $\gamma(G_l(t)) = v_l(t) \in \mathbb{R}^8$. \\

\begin{figure}[h!]
    \vspace{-5mm} \centering
    \begin{subfigure}{0.49\textwidth}
      \centering
      \includegraphics[width=\textwidth]{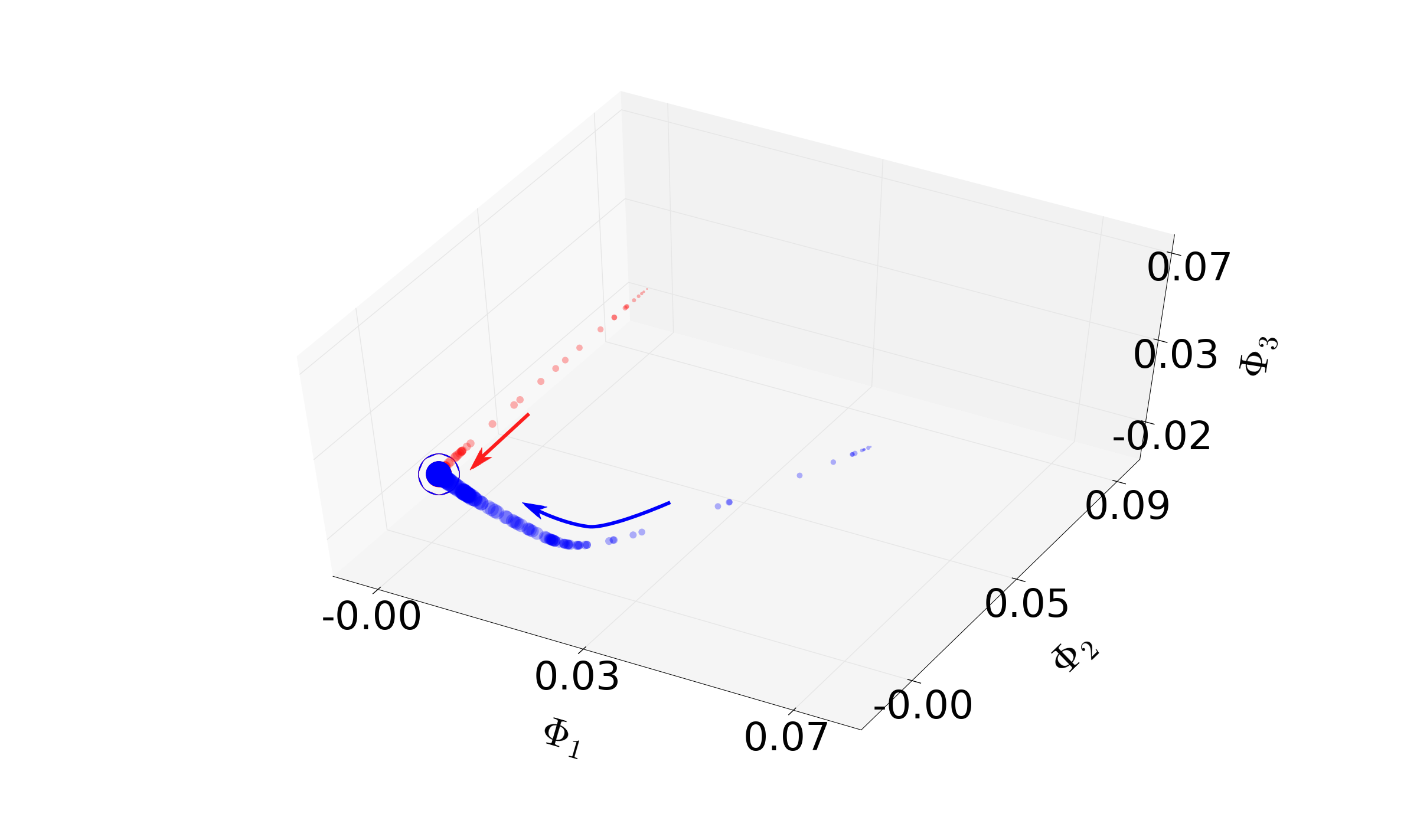}
      \subcaption{\label{fig:dmaps-results-regular}}
    \end{subfigure} %
    \begin{subfigure}{0.49\textwidth}
      \centering
      \includegraphics[width=\textwidth]{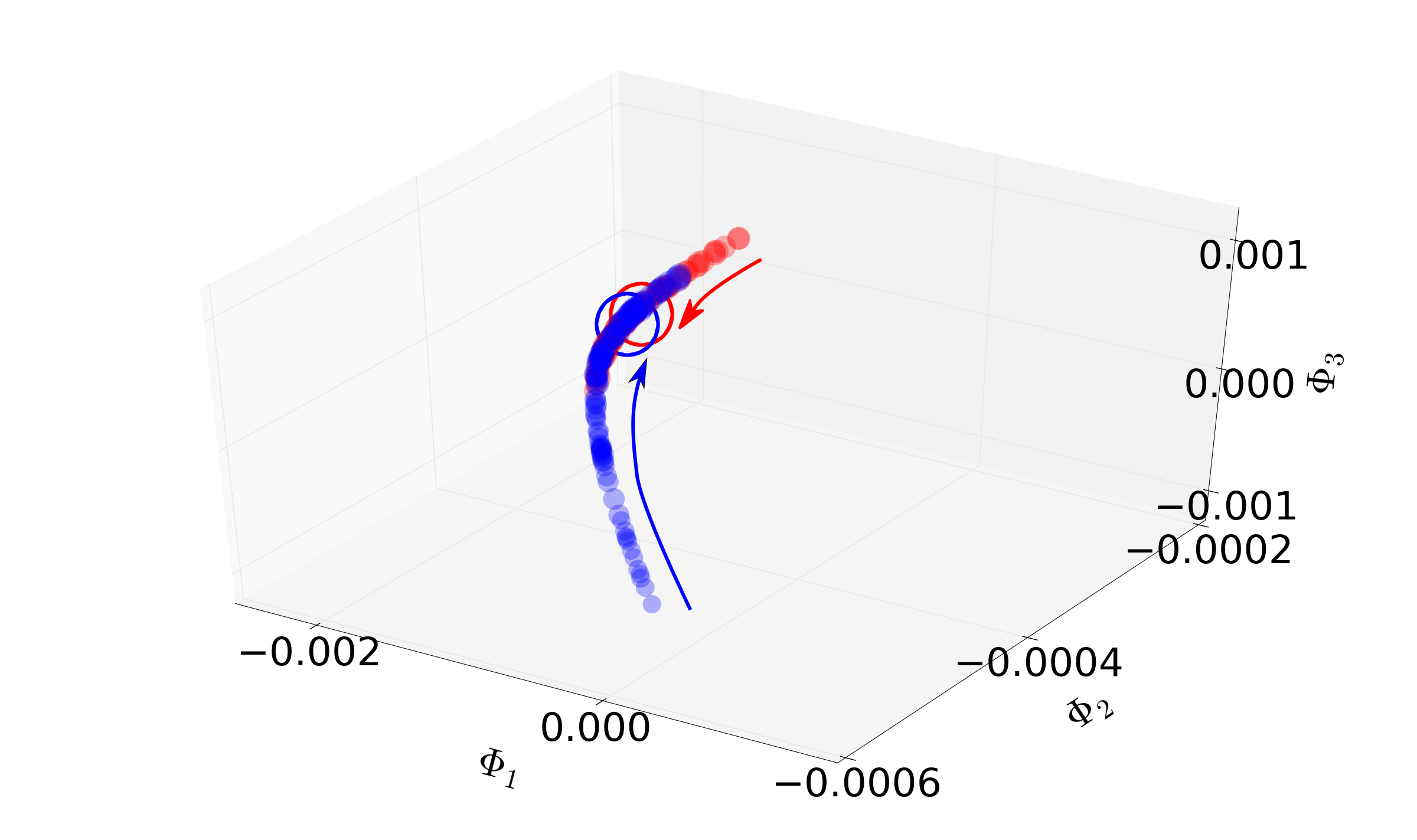}
      \subcaption{\label{fig:dmaps-results-zoom}}
    \end{subfigure}%
    \caption{DMAP of motif-based embeddings from two separate simulation
      trajectories: (a) embedding of two trajectories starting from
      different initial conditions. While the trajectories are initially
      mapped to separate regions of ``embedding space'', they eventually
      evolve to the same coarse stationary state.  Final states are
      circled, and point size grows as more steps are taken in each
      trajectory.  (b) Enlarged view of the final points in the previous
      plot, revealing the approach to a similar final state along
      opposite sides of the slowest eigenvector of its linearization.
      Here, the average of the final fifty points is circled. Due to the
      stochastic nature of the system, the final embeddings mildly
      fluctuate randomly about the shared stationary
      state. \label{fig:dmaps-results}}
  \end{figure}

  DMAPS was then applied to this set of $2N$ points, and the
  three-dimensional embedding using $\Phi_1$, $\Phi_2$ and $\Phi_3$ is
  shown in Fig. (\ref{fig:dmaps-results}).
  At first, the two trajectories are mapped to distant regions in
  $\mathbb{R}^3$ due to their different initial conditions.
  They evolve along different ``regions'' of the embedding, taking two
  distinct trajectories on their approach to the stationary state.
  Eventually, their embeddings meet as they arrive at this final, shared
  state, each asymptotically along opposite sides of the slowest
  eigenvector of the linearization of the steady state.
  DMAPS thus proves useful in elucidating both geometric and dynamic
  features of the system as shown in Figs. (\ref{fig:dmaps-rk}) and (\ref{fig:dmaps-results}). \\

  We see that both PCA and DMAPS, when combined with a suitable
  embedding of each graph, can help uncover useful information
  pertaining to the underlying dimensionality of the problem dynamics.

  \section{Conclusion}

  The equation free framework was successfully applied to the
  edge-conservative preferential attachment model, accelerating
  simulations through CPI and locating stationary states through coarse
  Newton-GMRES. This indicates potential avenues for improving
  simulation times of other complex network models, such as those
  found in epidemiology.
  Additionally, an underlying two-dimensional description was
  uncovered via PCA and DMAPS. These automated methods of
  dimensionality-reduction are quite general, and can in principle be
  applied in a wide range of network settings to probe hidden
  low-dimensional structure. For an application in the setting of
  labeled nodes, see ({\cite{kattis_modeling_2016}}). \\

  However, an open area of investigation is the interpretation of the output from
  the PCA and DMAPS techniques.
  As a linear map, PCA retains a clear relationship between its input
  and output, but is less effective when data lie on highly nonlinear
  manifolds.
  DMAPS may perform better the task of dimensionality reduction, but it
  is unclear how the embedding coordinates relate to physical features
  of the sampled networks.
  While this approach opens several promising avenues in coarse-graining
  complex network dynamics, it also reveals two ``bottlenecks'' for the
  process: (a) the selection of informative and practically computable
  graph distance metrics, necessary in the discovery of good coarse
  variables; and (b) the construction of network realizations consistent
  with (conditioned on) specific values of collective observables.
  Both items are, and we expect will continue being, the subject of
  intense investigation by many research groups, including ours.

  \begin{acknowledgement}
    The authors are grateful to Bal\'{a}zs R\'{a}th for helpful
    discussions and for several insightful suggestions that have greatly
    enhanced this work.
  \end{acknowledgement}

\end{onehalfspace}

\bibliographystyle{abbrv} \bibliography{multigraph-refs.bib}

\end{document}